\documentclass[journal]{IEEEtran}

\usepackage{etoolbox}
\robustify\bfseries
\usepackage{graphicx}
\usepackage{amsmath,amssymb,graphicx}
\usepackage[]{algorithm2e}
\usepackage{tabularx}
\usepackage{wasysym}
\usepackage{color,soul}
\usepackage{cite}
\usepackage{balance}
\usepackage{breqn}
\usepackage{makecell}
\usepackage{booktabs}
\usepackage{multirow}
\usepackage{siunitx}
\usepackage{xcolor}
\usepackage{hyperref}
\def\H{{\mathsf H}}
\def\T{{\mathsf T}}
\def\CC{{\mathbb C}}
\def\RR{{\mathbb R}}
\usepackage{caption}
\usepackage{arydshln}
\usepackage{enumitem}

\makeatletter
\newcommand*{\bigs}[1]{{\hbox{$\left#1\vbox to9\p@{}\right.\n@space$}}}
\makeatother

\usepackage{amssymb} %
\usepackage{pifont} %

\usepackage{flushend} %

\begin{document}

\title{Mixture to Mixture: Leveraging Close-talk Mixtures as Weak-supervision for Speech Separation}

\author{Zhong-Qiu Wang
\thanks{Manuscript received on Jan. 29, 2024; revised on May 19, 2024; accepted on Jun. 16, 2024.
}
\thanks{
Z.-Q. Wang is with the Department of Computer Science and Engineering at Southern University of Science and Technology, Shenzhen, Guangdong, China (e-mail: wang.zhongqiu41@gmail.com).}
}

\maketitle

\begin{abstract}
We propose \textit{mixture to mixture} (M2M) training, a weakly-supervised neural speech separation algorithm that leverages close-talk mixtures as a weak supervision for training discriminative models to separate far-field mixtures.
Our idea is that, for a target speaker, its close-talk mixture has a much higher signal-to-noise ratio (SNR) of the target speaker than any far-field mixtures, and hence could be utilized to design a weak supervision for separation.
To realize this, at each training step we feed a far-field mixture to a deep neural network (DNN) to produce an intermediate estimate for each speaker, and, for each of considered close-talk and far-field microphones, we linearly filter the DNN estimates and optimize a loss so that the filtered estimates of all the speakers can sum up to the mixture captured by each of the considered microphones.
Evaluation results on a $2$-speaker separation task in simulated reverberant conditions show that M2M can effectively leverage close-talk mixtures as a weak supervision for separating far-field mixtures.
\end{abstract}
\vspace{-0.1cm}

\IEEEpeerreviewmaketitle

\vspace{-0.2cm}
\section{Introduction}

\IEEEPARstart{D}{eep} learning has significantly elevated the performance of speech separation \cite{WDLreview} thanks to its strong modeling capabilities on human speech, especially since deep clustering \cite{Hershey2016} and permutation invariant training (PIT) \cite{Kolbak2017} solved the label permutation problem.
Modern neural speech separation models \cite{WDLreview, Hershey2016, Kolbak2017, Liu2019DeepCASA, Luo2019ConvTasNet, Maciejewski2020, Jenrungrot2020, Nachmani2020, Zeghidour2020, Tan2020, Wang2020chime, Wang2020css, ZhangJisi2020, Zhang2021ADL, Gu2021NeuralSpatialFilter, Luo2022GWF, Yoshioka2022VarArray, Wang2022GridNetjournal, Rixen2022, Tzinis2022Heterogeneous, Patterson2022Distance, Zmolikova2023SPM, Chetupalli2023EENDEDASeparation} are usually trained on simulated data in a supervised way, where clean speech is synthetically mixed with noise and competing speech in simulated reverberant rooms to generate paired clean and corrupted speech for supervised learning, where the clean speech can provide a \textit{sample-level} supervision for DNN training.
The trained models, however, often suffer from mismatches between simulated and real-recorded data, and are known to have severe generalization issues on real-recorded data \cite{Cornell2023, Aralikatti2022RAS, Leglaive2023CHiME7UDASE, Wisdom2020MixIT, Tzinis2022REMIXT, Wang2023UNSSOR}.

One way to address the problem is training models directly on real-recorded mixtures.
This however cannot be applied for supervised approaches since it is not possible to annotate the clean speech at each sample.
Another way is training unsupervised models on real-recorded mixtures, which usually makes strong assumptions on signal characteristics \cite{Wisdom2020MixIT, Bando2023NeuralFCAEUSIPO, Tzinis2022REMIXT, Fujimura2021, Wang2023UNSSOR}.
However, the performance could be fundamentally limited due to not leveraging any supervision and when the assumptions are not sufficiently satisfied in reality.

\begin{figure}
  \centering  
  \includegraphics[width=6.5cm]{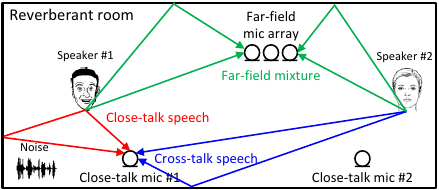}
  \vspace{-0.1cm}
  \caption{
  Illustration of task setup.
  Each close-talk mixture contains close- and cross-talk speech, and weak noises.
  Best viewed in color.
  }
  \label{physical_model_figure}
  \vspace{-0.6cm}
\end{figure}

While far-field mixtures are recorded in multi-speaker conditions, the close-talk mixture of each speaker is often recorded at the same time by using a close-talk microphone (e.g., in the AMI \cite{McCowan2006} and CHiME \cite{Barker2018CHiME5} setup).
See Fig. \ref{physical_model_figure} for an illustration.
The close-talk mixture of each speaker usually has a much higher SNR of the speaker than any far-field mixture.
Intuitively, it can be leveraged to train a model to increase the SNR of the speaker in far-field mixtures.
In this context, we propose to leverage close-talk mixtures as a weak supervision for separating far-field mixtures.
To realize this, we need to solve two major difficulties:
(a) close-talk mixtures are often not sufficiently clean, due to the contamination by cross-talk speech \cite{McCowan2006, Barker2018CHiME5, Watanabe2020CHiME6, Yu2022M2MeT};
and (b) close-talk mixtures are not time-aligned with far-field mixtures.
As a result, close-talk mixtures cannot be naively used as the training targets, and previous studies seldomly exploit them to build separation systems.

To overcome the two difficulties, we propose \textit{mixture to mixture} (M2M) training, where a DNN, taking in far-field mixtures as input, is discriminatively trained to produce an intermediate estimate for each target speaker in a way such that the intermediate estimates for all the speakers can be linearly filtered to recover the close-talk as well as far-field mixtures.
Following \cite{Wang2023UNSSOR}, the linear filters are computed via a neural forward filtering algorithm named forward convolutive prediction (FCP) \cite{Wang2021FCPjournal} based on the mixtures and intermediate DNN estimates.
We find that this linear filtering procedure can effectively address the above two difficulties.
This paper makes two major contributions:
\begin{itemize}[leftmargin=*,noitemsep,topsep=0pt]
\item We are the first seeking a way to leverage close-talk mixtures as a weak supervision for speech separation;
\item We propose a novel algorithm named M2M to exploit this weak supervision.
\end{itemize}
As an initial step, this paper evaluates M2M on a $2$-speaker separation task in simulated, reverberant conditions with weak noises.
The evaluation results show that M2M can effectively leverage the weak-supervision afforded by close-talk mixtures.

\vspace{-0.2cm}
\section{Related Work}\label{related_work}

There are several earlier studies on weakly-supervised separation.
In \cite{Stoller2018, Zhang2018WeaklySupervisedASS}, adversarially trained discriminators (in essence, source prior models) are used to encourage separation models to produce separation results with distributions similar to clean sources.
In \cite{Chang2019MIMOSpeech}, separation frontends are jointly trained with backend ASR models so that word transcriptions can be used to help frontends learn to separate.
In \cite{Pishdadian2020FindStrength}, a sound classifier is used to guide separation, by checking whether separated signals can be classified as target sound classes.
These approaches need clean sources, human annotations, and other models (e.g., discriminators, ASR models and sound classifiers).
Differently, M2M requires paired close-talk and far-field mixtures, which can be readily obtained during data collection by additionally using close-talk microphones, and it does not require other models.
On the other hand, close-talk mixtures exploited in M2M can provide a \textit{sample-level} supervision, which is much more fine-grained than source prior models, word transcriptions, and segment-level class labels.

\vspace{-0.15cm}
\section{Physical Model and Objectives}\label{proposed_physicalmodel}

In a reverberant environment with $C$ speakers (each wearing a close-talk microphone) and a far-field $P$-microphone array (see Fig.~\ref{physical_model_figure}), each recorded far-field and closed-talk mixture can be respectively formulated in the short-time Fourier transform (STFT) domain as follows:
\begin{align}
Y_p(t,f) &= \sum\nolimits_{c=1}^C X_p(c,t,f) + \varepsilon_p(t,f), \label{physical_model_ff} \\
Y_d(t,f) &= \sum\nolimits_{c=1}^C X_{d}(c,t,f) + \varepsilon_{d}(t,f), \label{physical_model_ct}
\end{align}
where $t$ indexes $T$ frames, $f$ indexes $F$ frequencies, $c$ indexes $C$ speakers, $d$ indexes $C$ close-talk microphones, and $p$ indexes $P$ far-field microphones. 
At time $t$ and frequency $f$, $Y_{p}(t,f)$, $X_p(c,t,f)$ and $\varepsilon_p(t,f)$ in (\ref{physical_model_ff}) respectively denote the far-field mixture, reverberant image of speaker $c$, and non-speech signals captured at far-field microphone $p$.
$Y_{d}(t,f)$, $X_{d}(c,t,f)$ and $\varepsilon_{d}(t,f)$ in (\ref{physical_model_ct}) respectively denote the STFT coefficients of the close-talk mixture, reverberant image of speaker $c$, and non-speech signals captured at close-talk microphone $d$ at time $t$ and frequency $f$.
In the rest of this paper, we refer to the corresponding spectrograms when dropping indices $p$, $c$, $d$, $t$ or $f$.
In this study, $\varepsilon$ is assumed a weak noise.

While speaker $c$ is talking, its close-talk speech $X_d(c)$, with $d=c$, in the close-talk mixture $Y_{d}$ is typically much stronger than cross-talk speech $X_d(c')$ by any other speaker $c'$ ($\neq c$).
By using close-talk mixtures as a weak supervision, we aim at training a DNN that can learn to estimate the reverberant speaker images (i.e., $X_p(c)$ for each speaker $c$ at a reference far-field microphone $p$), using only far-field mixtures as input.

\vspace{-0.15cm}
\section{M2M}\label{proposed_M2M}

\begin{figure}
  \centering  
  \includegraphics[width=8.5cm]{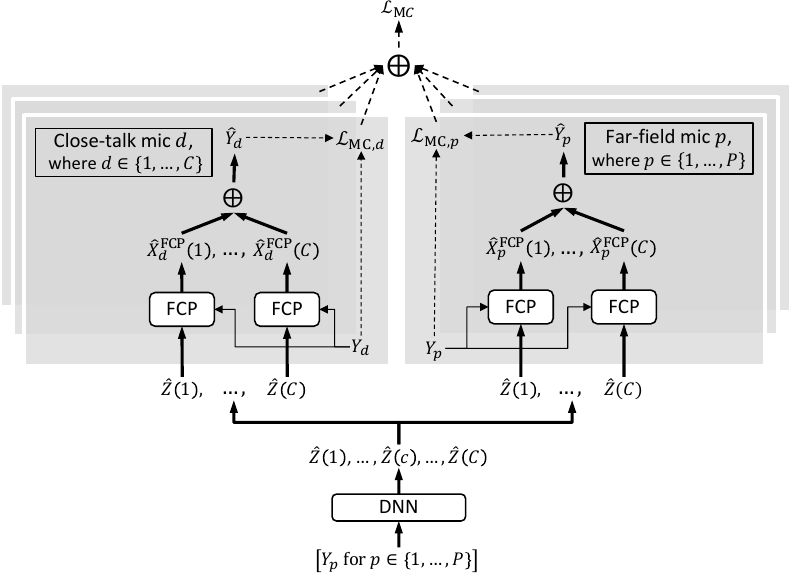}
  \vspace{-0.1cm}
  \caption{Illustration of M2M (described in first paragraph of \ref{proposed_M2M}).}
  \label{system_figure}
  \vspace{-0.6cm}
\end{figure}

Fig.~\ref{system_figure} illustrates M2M.
The DNN takes in far-field mixtures as input and produces an intermediate estimate $\hat{Z}(c)$ for each speaker $c$.
Each estimate $\hat{Z}(c)$ is then linearly filtered via FCP such that the filtered estimates can be summated to recover each of the close-talk and far-field mixtures.
This section describes the DNN setup, loss functions, and FCP filtering.

\vspace{-0.15cm}
\subsection{DNN Setup}

The DNN is trained to perform complex spectral mapping \cite{Tan2020, Wang2020chime, Wang2020css}, where the real and imaginary (RI) components of far-field mixtures are stacked as input features for the DNN to predict the RI components of $\hat{Z}(c)$ for each speaker $c$.
The DNN setup is described in \ref{experimental_setup} and the loss function in \ref{loss_description}.

\vspace{-0.15cm}
\subsection{Mixture-Constraint Loss}\label{loss_description}

We propose a mixture-constraint (MC) loss to encourage the DNN to produce an intermediate estimate $\hat{Z}$ that can be utilized to reconstruct the close-talk and far-field mixtures:
\begin{align}\label{loss_MC}
\mathcal{L}_{\text{MC}} = \sum\nolimits_{d=1}^C \mathcal{L}_{\text{MC},d} + \alpha \times \sum\nolimits_{p=1}^P \mathcal{L}_{\text{MC},p},
\end{align}
where $\mathcal{L}_{\text{MC},d}$ is the loss at close-talk microphone $d$, $\mathcal{L}_{\text{MC},p}$ at far-field microphone $p$, and $\alpha \in \RR_{>0}$ a weighting term.

$\mathcal{L}_{\text{MC},d}$ is defined, following the physical model in (\ref{physical_model_ct}), as
\begin{align}\label{loss_MC_close_talk_one}
\mathcal{L}_{}&{}_{\text{MC},d} = \sum\nolimits_{t,f} \mathcal{F} \Big( Y_{d}(t,f), \hat{Y}_{d}(t,f) \Big)  \nonumber \\
&= \sum\nolimits_{t,f} \mathcal{F} \Big( Y_{d}(t,f), \sum\nolimits_{c=1}^C \hat{X}_{d}^{\text{FCP}}(c,t,f) \Big)\nonumber \\
&= \sum\nolimits_{t,f} \mathcal{F} \Big( Y_{d}(t,f), \sum\nolimits_{c=1}^C \hat{\mathbf{g}}_{d}(c,f)^{\H}\ \widetilde{\hat{\mathbf{Z}}}(c,t,f) \Big),
\end{align}%
where $\widetilde{\hat{\mathbf{Z}}}(c,t,f)=[\hat{Z}(c,t-I,f),\dots,\hat{Z}(c,t+J,f)]^\T \in \CC^{I+1+J}$ stacks a window of T-F units, $\hat{\mathbf{g}}_{d}(c,f) \in \CC^{I+1+J}$ is a time-invariant FCP filter which will be detailed in Section \ref{FCP_description}, and $\mathcal{F}(\cdot, \cdot)$ is a distance measure to be described later.
In (\ref{loss_MC_close_talk_one}), the intermediate estimate $\hat{Z}(c)$ of each speaker $c$ is linearly filtered such that (a) the filtering result, $\hat{X}_{d}^{\text{FCP}}(c,t,f) = \hat{\mathbf{g}}_{d}(c,f)^{\H}\ \widetilde{\hat{\mathbf{Z}}}(c,t,f)$, can approximate $X_d(c)$, the cross-talk speech of speaker $c$ captured by close-talk microphone $d$; and (b) the filtering results of all the speakers can add up to the close-talk mixture $Y_d$ (i.e., $\hat{Y}_d=\sum\nolimits_{c=1}^C \hat{X}_{d}^{\text{FCP}}(c)$).
This way, we can leverage close-talk mixtures as a weak supervision for model training, and the linear filtering procedure can account for the mismatched time-alignment between close-talk and far-field mixtures.
Since the model is trained to reconstruct close-talk mixtures based on far-field mixtures, we name the algorithm \textit{mixture to mixture}.

$\mathcal{F}(\cdot, \cdot)$ in (\ref{loss_MC_close_talk_one}) computes a loss between the mixture $Y_d$ and reconstructed mixture $\hat{Y}_d$ based on the estimated RI components and their magnitude \cite{Wang2023UNSSOR}:
\begin{align}\label{L_D}
\resizebox{0.17\textwidth}{!}{$
\mathcal{F} \Big( Y_d(t,f), \hat{Y}_d(t,f) \Big) = 
$}
\resizebox{0.275\textwidth}{!}{$
\frac{\sum\nolimits_{\mathcal{O}\in \Omega} \big| \mathcal{O}(Y_d(t,f)) - \mathcal{O}(\hat{Y}_d(t,f)) \big| }{\sum\nolimits_{t',f'} \big|Y_d(t',f')\big|},
$}
\end{align}
where $\Omega=\{\mathcal{R},\mathcal{I},\mathcal{A}\}$ denotes a set of functions with $\mathcal{R}(\cdot)$ extracting the real part, $\mathcal{I}(\cdot)$ the imaginary part and $\mathcal{A}(\cdot)$ the magnitude of a complex number, $|\cdot|$ computes magnitude, and the denominator balances the losses at different microphones and across training mixtures.

Following (\ref{physical_model_ff}), $\mathcal{L}_{\text{MC},p}$ is similarly defined as follows:
\begin{align}\label{loss_MC_far_field}
\mathcal{L}_{}&{}_{\text{MC},p} = \sum\nolimits_{t,f} \mathcal{F} \Big( Y_p(t,f), \hat{Y}_p(t,f) \Big) \nonumber \\
&= \sum\nolimits_{t,f} \mathcal{F} \Big( Y_p(t,f), \sum\nolimits_{c=1}^C  \hat{X}_p^{\text{FCP}}(c,t,f) \Big) \nonumber \\
&= \sum\nolimits_{t,f} \mathcal{F} \Big( Y_p(t,f), \sum\nolimits_{c=1}^C \hat{\mathbf{g}}_p(c,f)^{\H}\ \widetilde{\hat{\mathbf{Z}}}(c,t,f) \Big),
\end{align}
where $\widetilde{\hat{\mathbf{Z}}}(c,t,f)=[\hat{Z}(c,t-M,f),\dots,\hat{Z}(c,t+N,f)]^\T \in \CC^{M+1+N}$ stacks a window of T-F units and $\hat{\mathbf{g}}_{p}(c,f) \in \CC^{M+1+N}$ is a time-invariant FCP filter to be described later.

We can configure the filter taps, $I,J,M$ and $N$, differently for close-talk and far-field microphones, considering that the microphones form a distributed rather than compact array.

\vspace{-0.15cm}
\subsection{FCP for Relative RIR Estimation}\label{FCP_description}

To compute $\mathcal{L}_{\text{MC}}$, the linear filters need to be first computed.
Each filter can be interpreted as the relative transfer function (RTF) relating the intermediate DNN estimate of a speaker to its reverberant image captured by another microphone.
Following \cite{Wang2023UNSSOR}, we employ FCP \cite{Wang2021FCPjournal} to estimate the RTFs.

Assuming that speakers do not move within each utterance, we estimate RTFs by solving the following problem:
\begin{align}\label{fcp_proj_mixture}
\resizebox{0.14\textwidth}{!}{$
\hat{\mathbf{g}}_r(c,f) =
\underset{\mathbf{g}_r(c,f)}{\text{argmin}}
$}
\sum\limits_t \frac{\Big| Y_r(t,f)-\mathbf{g}_r(c,f)^{\H}\ \widetilde{\hat{\mathbf{Z}}}(c,t,f) \Big|^2}{\hat{\lambda}_r(c,t,f)},
\end{align}
where the subscript $r$ indexes the $P$ far-field and $C$ close-talk microphones, and $\mathbf{g}_r(c,f)$ and $\widetilde{\hat{\mathbf{Z}}}(c,t,f)$ are defined in the text below (\ref{loss_MC_close_talk_one}) and (\ref{loss_MC_far_field}).
$\hat{\lambda}$ is a weighting term balancing the importance of each T-F unit.
For each close-talk microphone $d$ and speaker $c$, it is defined, following \cite{Wang2021FCPjournal}, as
\begin{align}\label{FCPweight_ct}
\hat{\lambda}_d(c,t,f) = \xi\times \text{max}(|Y_d|^2) + |Y_d(t,f)|^2,
\end{align}
where $\xi$ (set to $10^{-4}$) floors the weighting term and $\text{max}(\cdot)$ extracts the maximum value of a power spectrogram; and for each far-field microphone $p$ and speaker $c$, it is defined as 
\begin{align}\label{FCPweight_ff}
\hat{\lambda}_p(c,t,f) = \xi\times \text{max}(Q) + Q(t,f),
\end{align}
where $Q=\frac{1}{P}\sum_{p=1}^{P} |Y_{p}|^2$ averages the power spectrograms of far-field mixtures.
Notice that $\hat{\lambda}$ is computed differently for different microphones, as the energy level of each speaker is different at close-talk and far-field microphones.
Notice that (\ref{fcp_proj_mixture}) is a quadratic problem, where a closed-form solution can be readily computed.
We then plug $\hat{\mathbf{g}}_r(c,f)$ into (\ref{loss_MC_close_talk_one}) and (\ref{loss_MC_far_field}) to compute the loss, and train the DNN.

Although, in (\ref{fcp_proj_mixture}), $\hat{Z}(c)$ is linearly filtered to approximate $Y_r$, previous studies \cite{Wang2021FCPjournal, Wang2023UNSSOR} have suggested that the filtering result $\hat{\mathbf{g}}_r(c,f)^{\H}\widetilde{\hat{\mathbf{Z}}}(c,t,f)$ would approximate the speaker image $X_r(c,t,f)$, when $\hat{Z}(c)$ gets sufficiently accurate during training so that $\hat{Z}(c)$ becomes little correlated with sources other than $c$ (see detailed derivations in Appendix C of \cite{Wang2023UNSSOR}).
The estimated speaker image is named \textit{FCP-estimated image}:
\begin{align}\label{FCP_image}
\hat{X}_r^{\text{FCP}}(c,t,f) = \hat{\mathbf{g}}_r(c,f)^{\H}\ \widetilde{\hat{\mathbf{Z}}}(c,t,f).
\end{align}
The FCP-estimated images of all the speakers can be hence summated to reconstruct $Y_r$ in (\ref{loss_MC_close_talk_one}) and (\ref{loss_MC_far_field}).

At run time, we use the FCP-estimated image $\hat{X}_p^{\text{FCP}}(c)$ as the prediction for each speaker $c$ at a reference far-field microphone $p$.
We use the time-domain signal of the clean far-field image, $X_p(c)$, as the reference signal for evaluation.

\vspace{-0.15cm}
\subsection{Relations to, and Differences from, UNSSOR}

M2M is motivated by a recent algorithm named UNSSOR \cite{Wang2023UNSSOR}, an unsupervised neural speech separation algorithm designed for separating far-field mixtures.
UNSSOR is trained to optimize a loss similar to (\ref{loss_MC_far_field}), by leveraging the mixture signal at each microphone as a constraint to regularize DNN-estimated speaker images, and it can be successfully trained if the mixtures for training are over-determined (i.e., more microphones than sources) \cite{Wang2023UNSSOR}.
The major novelty of M2M is adapting UNSSOR for weakly-supervised separation by defining the MC loss not only on far-field microphones, but also on close-talk microphones to leverage the weak supervision afforded by close-talk mixtures to obtain better separation than UNSSOR, which is unsupervised.
In M2M, there are $C$ speakers, and $P$ far-field and $C$ close-talk microphones for loss computation.
The over-determined condition is hence naturally satisfied (i.e., $P+C > C$).
With that being said, only using the MC loss on close-talk mixtures (i.e., the first term in (\ref{loss_MC})) for training M2M would not lead to separation of speakers.
This is because, as is suggested in UNSSOR \cite{Wang2023UNSSOR}, the number of close-talk mixtures used for loss computation is not larger than the number of sources, and there would be an infinite number of DNN-estimated speaker images that can minimize the MC loss.
The second term in (\ref{loss_MC}) can help narrow down the infinite solutions to target speaker images.

\vspace{-0.2cm}
\section{Experimental Setup}\label{experimental_setup}

Since there are no earlier studies leveraging close-talk mixtures as a weak supervision for separation, to validate M2M we propose a simulated dataset so that clean reference signals can be used for evaluation.
Building upon the SMS-WSJ corpus \cite{Drude2019}, which only has far-field (FF) mixtures, we simulate SMS-WSJ-FF-CT, by adding close-talk (CT) mixtures.

\textbf{SMS-WSJ} \cite{Drude2019} is a popular corpus for $2$-speaker separation in reverberant conditions.
It has $33,561$ ($\sim$$87.4$ h), $982$ ($\sim$$2.5$ h) and $1,332$ ($\sim$$3.4$ h) $2$-speaker mixtures respectively for training, validation and testing.
The clean speech is sampled from the WSJ0 and WSJ1 corpus.
The simulated far-field array has $6$ microphones uniformly placed on a circle with a diameter of $20$ cm.
For each mixture, the speaker-to-array distance is drawn from the range $[1.0, 2.0]$ m, and the reverberation time (T60) from $[0.2, 0.5]$ s.
A white noise is added, at an energy level between the summation of the reverberant speech and the noise, sampled from the range $[20, 30]$ dB.
\textbf{SMS-WSJ-FF-CT} is created by adding a close-talk microphone for each speaker in each SMS-WSJ mixture.
The distance from each speaker to its close-talk microphone is uniformly sampled from the range $[10, 30]$ cm.
All the other setup for simulation remains the same.
This way, we can simulate the close-talk mixture of each speaker, and the far-field mixtures are exactly the same as the existing ones in SMS-WSJ.
The sampling rate is 8 kHz.

For STFT, the window size is $32$ ms and hop size $8$ ms.
TF-GridNet \cite{Wang2022GridNetjournal}, which has shown strong performance in major supervised speech separation benchmarks, is used as the DNN architecture.
Using the symbols defined in Table I of \cite{Wang2022GridNetjournal}, we set its hyper-parameters to $D=96$, $B=4$, $I=2$, $J=2$, $H=192$, $L=4$ and $E=4$ (please do not confuse these symbols with the ones in this paper).
We train it on $4$-second segments using a batch size of $4$.
The first far-field microphone is designated as the reference microphone.
We consider $6$-channel separation, where all the $6$ far-field microphones are used as input to M2M, and $1$-channel separation, where only the reference microphone signal can be used as input.
The evaluation metrics include signal-to-distortion ratio (SDR) \cite{Vincent2006a}, scale-invariant SDR (SI-SDR) \cite{LeRoux2019}, perceptual evaluation of speech quality (PESQ) \cite{Rix2001}, and extended short-time objective intelligibility (eSTOI) \cite{H.Taal2011}.

For comparison, we consider an unsupervised neural speech separation algorithm named UNSSOR \cite{Wang2023UNSSOR}, which is trained on far-field mixtures without using any supervision but also by optimizing a loss defined between linearly-filtered DNN estimates and observed mixtures.
In addition, we provide the results of PIT \cite{Kolbak2017}, trained in a supervised way assuming the availability of clean speaker images at far-field microphones, by using a loss defined, similarly to (\ref{L_D}), on the predicted real, imaginary and magnitude components.
Both baselines use the same TF-GridNet architecture and training setup as M2M, and their performance can be respectively viewed as the lower- and upper-bound performance of M2M.

\vspace{-0.2cm}
\section{Evaluation Results}

Table \ref{qulity_close_talk} presents the scores of close-talk mixtures, which are computed by using the close-talk speech of each speaker as reference and close-talk mixture as estimate.
We can see that the close-talk mixtures are not sufficiently clean (e.g., only $14.7$ dB in SI-SDR), due to the contamination by cross-talk speech, but the SNR of the target speaker is reasonably high.

Table \ref{m2m_results} reports the results of M2M when there are $P=6$ far-field microphones.
The reference signals for metric computation are the speaker images captured by the far-field reference microphone.
In row 1a-1e and 2a-2c, we tune the filter taps $I$, $J$, $M$ and $N$, and observe that the setup in 1a leads to the best separation.
Only one future tap (i.e., $J=1$ and $N=1$) is used in row 1a, and using more future taps are not helpful, likely because sound would travel $2.72 = 340 \times 0.008$ meters in $8$ ms (equal to the hop size of our system) if its speed in air is $340$ m/s, and this distance is already larger than the aperture size formed by the simulated close-talk and far-field microphones.
Compared to unsupervised UNSSOR in 4a, M2M in 1a produces clearly better separation; and compared with fully-supervised PIT in 4b, M2M in 1a shows competitive performance. These results indicate that M2M can effectively leverage the weak supervision afforded by close-talk mixtures

\begin{table}[]
\scriptsize
\centering
\sisetup{table-format=2.2,round-mode=places,round-precision=2,table-number-alignment = center,detect-weight=true,detect-inline-weight=math}
\caption{\textsc{Quality of Close-Talk Mixtures (Ref: Close-Talk Speech).}}
\vspace{-0.1cm}
\label{qulity_close_talk}
\setlength{\tabcolsep}{2pt}
\resizebox{0.7\columnwidth}{!}{
\begin{tabular}{
c %
S[table-format=1.1,round-precision=1] %
S[table-format=2.1,round-precision=1] %
S[table-format=1.2,round-precision=2] %
S[table-format=1.3,round-precision=3] %
}
\toprule
Dataset & {SI-SDR (dB)$\uparrow$} & {SDR (dB)$\uparrow$} & {PESQ$\uparrow$} & {eSTOI$\uparrow$} \\
\midrule
SMS-WSJ-FF-CT & 14.655501344786572 & 14.70267880020375 & 2.9178258743998526 & 0.8748214897000004 \\
\bottomrule
\end{tabular}%
}

\vspace{0.3cm}

\scriptsize
\centering
\sisetup{table-format=2.2,round-mode=places,round-precision=2,table-number-alignment = center,detect-weight=true,detect-inline-weight=math}
\caption{\textsc{Separation Results on SMS-WSJ-FF-CT ($P=6$)\\(Reference: Speaker Images at Far-Field Reference Mic)}}
\vspace{-0.1cm}
\label{m2m_results}
\setlength{\tabcolsep}{2pt}
\resizebox{\columnwidth}{!}{
\begin{tabular}{
r %
c %
c %
S[table-format=1,round-precision=0] %
c %
S[table-format=2.1,round-precision=1]
S[table-format=2.1,round-precision=1]
S[table-format=1.2,round-precision=2]
S[table-format=1.3,round-precision=3]
}
\toprule
{\multirow{2}{*}{\rotatebox[origin=c]{0}{Row}}} & {\multirow{2}{*}{Type}} & {\multirow{2}{*}{Systems}} & {\multirow{2}{*}{$I/J/M/N$}} & {\multirow{2}{*}{$\alpha$}} & {SI-SDR} & {SDR} & {\multirow{2}{*}{PESQ$\uparrow$}} & {\multirow{2}{*}{eSTOI$\uparrow$}} \\
 & & & & & {(dB)$\uparrow$} & {(dB)$\uparrow$} & & \\
\midrule
0 & {-} & Mixture & {-} & {-} & -0.03418624199922475 & 0.05634484792852443 & 1.8676067904846088 & 0.6029869126586817 \\
\midrule
1a & {weakly-sup.} & M2M & {$19/1/19/1$} & $1.0$ & 16.916469185537583 & 17.91053948728185 & 3.8504215581370547 & 0.9314890807728113 \\
1b & {weakly-sup.} & M2M & {$19/2/19/1$} & $1.0$ & 15.821417227856585 & 16.729973250338993 & 3.7608183256767176 & 0.9155952393316257 \\
1c & {weakly-sup.} & M2M & {$19/3/19/1$} & $1.0$ & 16.25633160071867 & 17.209361877800006 & 3.7874937572994747 & 0.9237856837361546 \\
1d & {weakly-sup.} & M2M & {$19/4/19/1$} & $1.0$ & 15.903067430698064 & 16.8148912301099 & 3.744766241258329 & 0.9167091857866456 \\
1e & {weakly-sup.} & M2M & {$19/5/19/1$} & $1.0$ & 15.81721593214719 & 16.72294154993484 & 3.7535114190778933 & 0.9141779621799273 \\
\midrule
2a & {weakly-sup.} & M2M & {$19/0/19/0$} & $1.0$ & 15.995087914949131 & 16.916007069501042 & 3.7723257024366936 & 0.9180873799539672 \\
2b & {weakly-sup.} & M2M & {$19/2/19/2$} & $1.0$ & 15.836690290811452 & 16.76819917271272 & 3.7402943178966597 & 0.9138903977642878 \\
2c & {weakly-sup.} & M2M & {$19/3/19/3$} & $1.0$ & 16.14228533404703 & 17.07691504071085 & 3.7692988590077237 & 0.9210632224327454 \\
\midrule
4a & Unsupervised & UNSSOR \cite{Wang2023UNSSOR} & {-} & {-} & 14.683185558306059 & 15.587776263552826 & 3.4448701904879675 & 0.8859511559809052 \\
4b & Supervised & PIT \cite{Kolbak2017} & {-} & {-} & 18.865480476196062 & 19.361938623387395 & 4.061564921214058 & 0.9497483915389929 \\
\bottomrule
\end{tabular}%
}

\vspace{0.3cm}

\scriptsize
\centering
\sisetup{table-format=2.2,round-mode=places,round-precision=2,table-number-alignment = center,detect-weight=true,detect-inline-weight=math}
\caption{\textsc{Separation Results on SMS-WSJ-FF-CT ($P=1$)\\(Reference: Speaker Images at Far-Field Reference Mic)}}
\vspace{-0.1cm}
\label{m2m_1ch_results}
\setlength{\tabcolsep}{2pt}
\resizebox{1.0\columnwidth}{!}{
\begin{tabular}{
r %
c %
c %
S[table-format=1,round-precision=0] %
c %
S[table-format=2.1,round-precision=1]
S[table-format=2.1,round-precision=1]
S[table-format=1.2,round-precision=2]
S[table-format=1.3,round-precision=3]
}
\toprule
{\multirow{2}{*}{\rotatebox[origin=c]{0}{Row}}} & {\multirow{2}{*}{Type}} & {\multirow{2}{*}{Systems}} & {\multirow{2}{*}{$I/J/M/N$}} & {\multirow{2}{*}{$\alpha$}} & {SI-SDR} & {SDR} & {\multirow{2}{*}{PESQ$\uparrow$}} & {\multirow{2}{*}{eSTOI$\uparrow$}} \\
 & & & & & {(dB)$\uparrow$} & {(dB)$\uparrow$} & & \\
\midrule
0 & {-} & Mixture & {-} & {-} & -0.03418624199922475 & 0.05634484792852443 & 1.8676067904846088 & 0.6029869126586817 \\
\midrule
1 & {weakly-sup.} & M2M & {$19/1/19/1$} & $1.0$ & -2.552021274096078 & -1.2544930481602026 & 1.6417852955627013 & 0.4793260629175127 \\
\midrule
2a & {weakly-sup.} & M2M & {$19/1/19/1$} & $1/5$ & 12.019405789203471 & 12.92972111437815 & 3.4105769948021427 & 0.856535483047431 \\
2b & {weakly-sup.} & M2M & {$19/1/19/1$} & $1/6$ & 11.810385831617737 & 12.686472606466173 & 3.3986065413679802 & 0.8527719136210271 \\
2c & {weakly-sup.} & M2M & {$19/1/19/1$} & $1/7$ & 12.73210758511844 & 13.669340168596634 & 3.4999145123514683 & 0.871918245198691 \\
2d & {weakly-sup.} & M2M & {$19/1/19/1$} & $1/8$ & 12.113955689401173 & 13.004764606002007 & 3.4475293639185907 & 0.8593423238809351 \\
\midrule
3 & Supervised & PIT \cite{Kolbak2017} & {-} & {-} & 13.676703066506365 & 14.127530009501628 & 3.612348065719948 & 0.8839671653012012 \\
\bottomrule\vspace{-0.7cm}
\end{tabular}
}
\end{table}

Table \ref{m2m_1ch_results} reports the results when $P=1$, where M2M only takes in the far-field mixture signal at the reference microphone as input and is trained to reconstruct the input mixture and close-talk mixtures.
When the weight $\alpha$ in (\ref{loss_MC}) is $1.0$, the DNN could just copy its input as the output (e.g., $\hat{Z}(c)=Y_1$) to optimize the loss on the far-field mixture (i.e., the second term in (\ref{loss_MC})) to zero, causing the loss on close-talk mixtures not optimized well.
To avoid this, we apply a smaller weight $\alpha$ to the loss on the far-field mixture so that the DNN can focus on reconstructing close-talk mixtures.
From row 1 and 2a-2d of Table \ref{m2m_1ch_results}, we can see that this strategy works, and M2M obtains competitive results compared to monaural supervised PIT in row 3.

In our experiments, we observe that, even if FCP is performed in each frequency independently from the others, M2M does not suffer from the frequency permutation problem \cite{Gannot2017, Sawada2019BSSReview}, which needs to be carefully dealt with in UNSSOR \cite{Wang2023UNSSOR} and in many frequency-domain blind source separation algorithms \cite{Sawada2019BSSReview}.
This is possibly because each close-talk mixture has a high SNR of the target speaker, which can give a hint to M2M regarding what the target source is across all the frequencies of each output spectrogram.

A sound demo based on the experiments is provided in this link \url{https://zqwang7.github.io/demos/M2M_demo/index.html}.

\vspace{-0.2cm}
\section{Conclusion}

We have proposed M2M, which leverages close-talk mixtures as a weak supervision for training neural speech separation models to separate far-field mixtures.
Evaluation results on $2$-speaker separation in simulated conditions show the effectiveness of M2M.
Future research will modify and evaluate M2M on real-recorded far-field and close-talk mixtures.

In closing, the key scientific contribution of this paper, we emphasize, is a novel methodology that directly trains neural source separation models based on paired mixtures, where the higher-SNR mixture can serve as a weak supervision for separating the lower-SNR mixture.
This concept of \textit{mixture-to-mixture training}, we believe, would motivate the design of many algorithms in future research in neural source separation.

\bibliographystyle{IEEEtran}
\bibliography{references.bib}

\begin{thebibliography}{10}
\providecommand{\url}[1]{#1}
\csname url@samestyle\endcsname
\providecommand{\newblock}{\relax}
\providecommand{\bibinfo}[2]{#2}
\providecommand{\BIBentrySTDinterwordspacing}{\spaceskip=0pt\relax}
\providecommand{\BIBentryALTinterwordstretchfactor}{4}
\providecommand{\BIBentryALTinterwordspacing}{\spaceskip=\fontdimen2\font plus
\BIBentryALTinterwordstretchfactor\fontdimen3\font minus
  \fontdimen4\font\relax}
\providecommand{\BIBforeignlanguage}[2]{{%
\expandafter\ifx\csname l@#1\endcsname\relax
\typeout{** WARNING: IEEEtran.bst: No hyphenation pattern has been}%
\typeout{** loaded for the language `#1'. Using the pattern for}%
\typeout{** the default language instead.}%
\else
\language=\csname l@#1\endcsname
\fi
#2}}
\providecommand{\BIBdecl}{\relax}
\BIBdecl

\bibitem{WDLreview}
D.~Wang and J.~Chen, ``{Supervised Speech Separation Based on Deep Learning: An
  Overview},'' \emph{IEEE/ACM Trans. Audio, Speech, Lang. Process.}, vol.~26,
  no.~10, pp. 1702--1726, 2018.

\bibitem{Hershey2016}
J.~R. Hershey, Z.~Chen, J.~{Le Roux}, and S.~Watanabe, ``{Deep Clustering:
  Discriminative Embeddings for Segmentation and Separation},'' in
  \emph{ICASSP}, 2016, pp. 31--35.

\bibitem{Kolbak2017}
M.~Kolb{\ae}k, D.~Yu, Z.-H. Tan, and J.~Jensen, ``{Multitalker Speech
  Separation with Utterance-Level Permutation Invariant Training of Deep
  Recurrent Neural Networks},'' \emph{IEEE/ACM Trans. Audio, Speech, Lang.
  Process.}, vol.~25, no.~10, pp. 1901--1913, 2017.

\bibitem{Liu2019DeepCASA}
Y.~Liu and D.~Wang, ``{Divide and Conquer: A Deep CASA Approach to
  Talker-Independent Monaural Speaker Separation},'' \emph{IEEE/ACM Trans.
  Audio, Speech, Lang. Process.}, vol.~27, no.~12, pp. 2092--2102, 2019.

\bibitem{Luo2019ConvTasNet}
Y.~Luo and N.~Mesgarani, ``{Conv-TasNet: Surpassing Ideal Time-Frequency
  Magnitude Masking for Speech Separation},'' \emph{IEEE/ACM Trans. Audio,
  Speech, Lang. Process.}, vol.~27, pp. 1256--1266, 2019.

\bibitem{Maciejewski2020}
M.~MacIejewski, G.~Wichern, E.~McQuinn, and J.~{Le Roux}, ``{WHAMR!: Noisy and
  Reverberant Single-Channel Speech Separation},'' in \emph{ICASSP}, 2020, pp.
  696--700.

\bibitem{Jenrungrot2020}
T.~Jenrungrot, V.~Jayaram \emph{et~al.}, ``{The Cone of Silence: Speech
  Separation by Localization},'' in \emph{NeurIPS}, 2020.

\bibitem{Nachmani2020}
E.~Nachmani, Y.~Adi, and L.~Wolf, ``{Voice Separation with An Unknown Number of
  Multiple Speakers},'' in \emph{ICML}, 2020, pp. 7121--7132.

\bibitem{Zeghidour2020}
N.~Zeghidour and D.~Grangier, ``{Wavesplit: End-to-End Speech Separation by
  Speaker Clustering},'' \emph{IEEE/ACM Trans. Audio, Speech, Lang. Process.},
  vol.~29, pp. 2840--2849, 2021.

\bibitem{Tan2020}
K.~Tan and D.~Wang, ``{Learning Complex Spectral Mapping With Gated
  Convolutional Recurrent Networks for Monaural Speech Enhancement},''
  \emph{IEEE/ACM Trans. Audio, Speech, Lang. Process.}, vol.~28, pp. 380--390,
  2020.

\bibitem{Wang2020chime}
Z.-Q. Wang, P.~Wang, and D.~Wang, ``{Complex Spectral Mapping for Single- and
  Multi-Channel Speech Enhancement and Robust ASR},'' \emph{IEEE/ACM Trans.
  Audio, Speech, Lang. Process.}, vol.~28, pp. 1778--1787, 2020.

\bibitem{Wang2020css}
------, ``{Multi-Microphone Complex Spectral Mapping for Utterance-Wise and
  Continuous Speech Separation},'' \emph{IEEE/ACM Trans. Audio, Speech, Lang.
  Process.}, vol.~29, pp. 2001--2014, 2021.

\bibitem{ZhangJisi2020}
J.~Zhang, C.~Zorila, R.~Doddipatla, and J.~Barker, ``{On End-to-End
  Multi-Channel Time Domain Speech Separation in Reverberant Environments},''
  in \emph{ICASSP}, 2020, pp. 6389--6393.

\bibitem{Zhang2021ADL}
Z.~Zhang, Y.~Xu, M.~Yu, S.~X. Zhang, L.~Chen, D.~{S. Williamson}, and D.~Yu,
  ``{Multi-Channel Multi-Frame ADL-MVDR for Target Speech Separation},''
  \emph{IEEE/ACM Trans. Audio, Speech, Lang. Process.}, vol.~29, pp.
  3526--3540, 2021.

\bibitem{Gu2021NeuralSpatialFilter}
R.~Gu, S.~X. Zhang, Y.~Zou, and D.~Yu, ``{Complex Neural Spatial Filter:
  Enhancing Multi-Channel Target Speech Separation in Complex Domain},''
  \emph{IEEE Signal Process. Lett.}, vol.~28, pp. 1370--1374, 2021.

\bibitem{Luo2022GWF}
Y.~Luo, ``{A Time-domain Generalized Wiener Filter for Multi-channel Speech
  Separation},'' \emph{IEEE/ACM Trans. Audio, Speech, Lang. Process.}, vol.~30,
  pp. 3008 -- 3019, 2022.

\bibitem{Yoshioka2022VarArray}
T.~Yoshioka, X.~Wang, D.~Wang, M.~Tang, Z.~Zhu, Z.~Chen, and N.~Kanda,
  ``{VarArray: Array-Geometry-Agnostic Continuous Speech Separation},'' in
  \emph{ICASSP}, 2022.

\bibitem{Wang2022GridNetjournal}
Z.-Q. Wang, S.~Cornell, S.~Choi, Y.~Lee, B.-Y. Kim, and S.~Watanabe,
  ``{TF-GridNet: Integrating Full- and Sub-Band Modeling for Speech
  Separation},'' \emph{IEEE/ACM Transactions on Audio, Speech, and Language
  Processing}, vol.~31, pp. 3221--3236, 2023.

\bibitem{Rixen2022}
J.~Rixen and M.~Renz, ``{SFSRNet: Super-Resolution for Single-Channel Audio
  Source Separation},'' in \emph{AAAI}, 2022.

\bibitem{Tzinis2022Heterogeneous}
E.~Tzinis, G.~Wichern, A.~Subramanian, P.~Smaragdis, and J.~{Le Roux},
  ``{Heterogeneous Target Speech Separation},'' in \emph{Interspeech}, 2022,
  pp. 1796--1800.

\bibitem{Patterson2022Distance}
K.~Patterson, K.~Wilson, S.~Wisdom, and J.~{R. Hershey}, ``{Distance-Based
  Sound Separation},'' in \emph{Interspeech}, 2022, pp. 901--905.

\bibitem{Zmolikova2023SPM}
K.~Zmolikova, M.~Delcroix, T.~Ochiai, K.~Kinoshita, J.~Cernocky, and D.~Yu,
  ``{Neural Target Speech Extraction: An overview},'' \emph{IEEE Signal
  Processing Magazine}, vol.~40, no.~3, pp. 8--29, 2023.

\bibitem{Chetupalli2023EENDEDASeparation}
S.~R. Chetupalli and E.~A. Habets, ``{Speaker Counting and Separation From
  Single-Channel Noisy Mixtures},'' \emph{IEEE/ACM Trans. Audio, Speech, Lang.
  Process.}, vol.~31, pp. 1681--1692, 2023.

\bibitem{Cornell2023}
S.~Cornell, M.~Wiesner, S.~Watanabe, D.~Raj, X.~Chang, P.~Garcia, Y.~Masuyama,
  Z.-Q. Wang, S.~Squartini \emph{et~al.}, ``{The CHiME-7 DASR Challenge:
  Distant Meeting Transcription with Multiple Devices in Diverse Scenarios},''
  in \emph{CHiME}, 2023.

\bibitem{Aralikatti2022RAS}
R.~Aralikatti, C.~Boeddeker, G.~Wichern, A.~S. Subramanian \emph{et~al.},
  ``{Reverberation as Supervision for Speech Separation},'' in \emph{ICASSP},
  2023.

\bibitem{Leglaive2023CHiME7UDASE}
S.~Leglaive, L.~Borne, E.~Tzinis, M.~Sadeghi, M.~Fraticelli, S.~Wisdom,
  M.~Pariente, D.~Pressnitzer, and J.~{R. Hershey}, ``{The CHiME-7 UDASE Task:
  Unsupervised Domain Adaptation for Conversational Speech Enhancement},'' in
  \emph{CHiME}, 2023.

\bibitem{Wisdom2020MixIT}
S.~Wisdom, E.~Tzinis, H.~Erdogan, R.~J. Weiss, K.~Wilson, and J.~R. Hershey,
  ``{Unsupervised Sound Separation using Mixture Invariant Training},'' in
  \emph{NeurIPS}, 2020.

\bibitem{Tzinis2022REMIXT}
E.~Tzinis, Y.~Adi, V.~K. Ithapu, B.~Xu, P.~Smaragdis, and A.~Kumar, ``{RemixIT:
  Continual Self-Training of Speech Enhancement Models via Bootstrapped
  Remixing},'' \emph{IEEE J. of Selected Topics in Signal Process.}, vol.~16,
  no.~6, pp. 1329--1341, 2022.

\bibitem{Wang2023UNSSOR}
Z.-Q. Wang and S.~Watanabe, ``{UNSSOR: Unsupervised Neural Speech Separation by
  Leveraging Over-determined Training Mixtures},'' in \emph{NeurIPS}, 2023.

\bibitem{Bando2023NeuralFCAEUSIPO}
Y.~Bando, Y.~Masuyama, A.~A. Nugraha, and K.~Yoshii, ``{Neural Fast Full-Rank
  Spatial Covariance Analysis for Blind Source Separation},'' in
  \emph{EUSIPCO}, 2023.

\bibitem{Fujimura2021}
T.~Fujimura, Y.~Koizumi, K.~Yatabe, and R.~Miyazaki, ``{Noisy-target Training:
  A Training Strategy for DNN-based Speech Enhancement without Clean Speech},''
  in \emph{EUSIPCO}, 2021, pp. 436--440.

\bibitem{McCowan2006}
J.~Carletta, S.~Ashby, S.~Bourban, M.~Flynn \emph{et~al.}, ``{The AMI Meeting
  Corpus: A Pre-Announcement},'' in \emph{Machine Learning for Multimodal
  Interaction}, vol. 3869, 2006, pp. 28--39.

\bibitem{Barker2018CHiME5}
J.~Barker, S.~Watanabe, E.~Vincent, and J.~Trmal, ``{The Fifth 'CHiME' Speech
  Separation and Recognition Challenge: Dataset, Task and Baselines},'' in
  \emph{Interspeech}, 2018, pp. 1561--1565.

\bibitem{Watanabe2020CHiME6}
S.~Watanabe, M.~Mandel, J.~Barker, E.~Vincent \emph{et~al.}, ``{CHiME-6
  Challenge: Tackling Multispeaker Speech Recognition for Unsegmented
  Recordings},'' in \emph{arXiv preprint arXiv:2004.09249}, 2020.

\bibitem{Yu2022M2MeT}
F.~Yu, S.~Zhang, Y.~Fu, L.~Xie, S.~Zheng, Z.~Du, W.~Huang, P.~Guo
  \emph{et~al.}, ``{M2MeT: The ICASSP 2022 Multi-Channel Multi-Party Meeting
  Transcription Challenge},'' in \emph{ICASSP}, 2022, pp. 6167--6171.

\bibitem{Wang2021FCPjournal}
Z.-Q. Wang, G.~Wichern, and J.~{Le Roux}, ``{Convolutive Prediction for
  Monaural Speech Dereverberation and Noisy-Reverberant Speaker Separation},''
  \emph{IEEE/ACM Trans. Audio, Speech, Lang. Process.}, vol.~29, pp.
  3476--3490, 2021.

\bibitem{Stoller2018}
D.~Stoller, S.~Ewert, and S.~Dixon, ``{Adversarial Semi-Supervised Audio Source
  Separation Applied to Singing Voice Extraction},'' in \emph{ICASSP}, 2018,
  pp. 2391--2395.

\bibitem{Zhang2018WeaklySupervisedASS}
N.~Zhang, J.~Yan, and Y.~Zhou, ``{Weakly Supervised Audio Source Separation via
  Spectrum Energy Preserved Wasserstein Learning},'' in \emph{IJCAI}, 2018, pp.
  4574--4580.

\bibitem{Chang2019MIMOSpeech}
X.~Chang, W.~Zhang, Y.~Qian, J.~{Le Roux}, and S.~Watanabe, ``{MIMO-SPEECH:
  End-to-End Multi-Channel Multi-Speaker Speech Recognition},'' in \emph{ASRU},
  2019, pp. 237--244.

\bibitem{Pishdadian2020FindStrength}
F.~Pishdadian, G.~Wichern, and J.~{Le Roux}, ``{Finding Strength in Weakness:
  Learning to Separate Sounds with Weak Supervision},'' in \emph{IEEE/ACM
  Trans. Audio, Speech, Lang. Process.}, vol.~28, 2020, pp. 2386--2399.

\bibitem{Drude2019}
L.~Drude, J.~Heitkaemper, C.~Boeddeker, and R.~Haeb-Umbach, ``{SMS-WSJ:
  Database, Performance Measures, and Baseline Recipe for Multi-Channel Source
  Separation and Recognition},'' in \emph{arXiv preprint arXiv:1910.13934},
  2019.

\bibitem{Vincent2006a}
E.~Vincent, R.~Gribonval, and C.~F{\'{e}}votte, ``{Performance Measurement in
  Blind Audio Source Separation},'' \emph{IEEE Trans. Audio, Speech, Lang.
  Process.}, vol.~14, no.~4, pp. 1462--1469, 2006.

\bibitem{LeRoux2019}
J.~{Le Roux}, S.~Wisdom, H.~Erdogan, and J.~R. {Hershey}, ``{SDR - Half-Baked
  or Well Done?}'' in \emph{ICASSP}, 2019, pp. 626--630.

\bibitem{Rix2001}
A.~Rix, J.~Beerends \emph{et~al.}, ``{Perceptual Evaluation of Speech Quality
  (PESQ)-A New Method for Speech Quality Assessment of Telephone Networks and
  Codecs},'' in \emph{ICASSP}, vol.~2, 2001, pp. 749--752.

\bibitem{H.Taal2011}
C.~{H. Taal}, R.~{C. Hendriks} \emph{et~al.}, ``{An Algorithm for
  Intelligibility Prediction of Time-Frequency Weighted Noisy Speech},''
  \emph{IEEE Trans. Audio, Speech, Lang. Process.}, vol.~19, no.~7, pp.
  2125--2136, 2011.

\bibitem{Gannot2017}
S.~Gannot, E.~Vincent \emph{et~al.}, ``{A Consolidated Perspective on
  Multi-Microphone Speech Enhancement and Source Separation},'' \emph{IEEE/ACM
  Trans. Audio, Speech, Lang. Process.}, vol.~25, pp. 692--730, 2017.

\bibitem{Sawada2019BSSReview}
H.~Sawada, N.~Ono, H.~Kameoka, D.~Kitamura, and H.~Saruwatari, ``{A Review of
  Blind Source Separation Methods: Two Converging Routes to ILRMA Originating
  from ICA and NMF},'' \emph{APSIPA Trans. on Signal and Info. Process.},
  vol.~8, pp. 1--14, 2019.

\end{thebibliography}

\end{document}